\documentclass{appolb}
\usepackage{graphicx}

\begin{document}
\title{High-resolution Photoemission Study of Cd$_{2}$Re$_{2}$O$_{7}$%
\thanks{Presented at the Strongly Correlated Electron Systems 
Conference, Krak\'ow 2002}%
}


\author{A. Irizawa, A. Higashiya, S. Kasai, T. Sasabayashi, A. Shigemoto, A. Sekiyama, S. Imada, S. Suga
\address{Division of Materials Physics, Graduate School of Engineering Science, Osaka University, Osaka 560-8531, Japan}
\and
H. Sakai, H. Ohno, M. Kato, K. Yoshimura
\address{Department of Chemistry, Graduate School of Science, Kyoto University, Kyoto 606-8502, Japan}
}
\maketitle


\begin{abstract}
High-resolution bulk-sensitive photoemission has been measured for pyrochlore oxide Cd$_{2}$Re$_{2}$O$_{7}$. Temperature variations of the spectral shapes between 20 and 250 K are obviously observed for the Re 4f inner core and the valence band spectra. The Re 5d states are dominant for the DOS near the Fermi level.
\end{abstract}


  
\section{Introduction}
Recently, Cd$_{2}$Re$_{2}$O$_{7}$ has been found to be a superconductor with 
Tc$\sim $1 K~\cite{cit1,cit2}. This is the first superconductor found among the large 
family of pyrochlore oxides with the formula of A$_{2}$B$_{2}$O$_{7}$ 
(A=rare earth or late transition metals, B=transition metals). In this 
structure, A and B cations are 4- and 6-cordinated by oxygen anions. The 
A-O$_{4}$ tetrahedra are connected as forming a pyrochlore lattice with 
straight A-O-A bonds, while the B-O$_{6}$ octahedra form a pyrochlore 
lattice with the bent B-O-B bonds with the angle of $110$$\sim $$140^\circ$. Assuming 
the electronic configurations in Cd$_{2}$Re$_{2}$O$_{7}$ as formally Cd$^{2 
+ }$ 4d$^{10}$ and Re$^{5 + }$ 4f$^{14}$5d$^{2}$, the electronic and 
magnetic properties are primarily dominated by the Re 5d electrons. 
Cd$_{2}$Re$_{2}$O$_{7}$ shows an anomaly at 200 K in the electric 
resistivity, magnetic susceptibility, specific heat and Hall coefficient~
\cite{cit1,cit2,cit3}. An X-ray diffraction measurement reveals the existence of a 
second-order structural phase transition at this temperature with changing 
its symmetry from an ideal cubic \textit{Fd}3$m$ to a lower symmetric cubic $F$43$m$ with 
lowering temperature~\cite{cit4}. The resistivity is almost temperature independent 
near the room temperature and drops abruptly below 200 K. The magnetic 
susceptibility decreases also below 200 K while it shows weak temperature 
dependence at higher temperature with a broad maximum near 290 K. Although 
the origin of this transition is not clear, it induces a large change in 
electronic properties. Furthermore, another transition is shown not only in 
electric resistivity with a hysteresis at around 120 K, but also in specific 
heat and thermoelectric power in the form of kink, whereas magnetic 
susceptibility shows no change~\cite{cit5}. The magnetoresistance goes once to zero 
with increasing temperature toward 120 K from low temperatures, but it is 
revived between 120 and 200 K. This transition near 120 K is also pointed 
out to be due to a possible electronic structural change at the Fermi level.

In order to directly study the electronic states, a high-resolution 
bulk-sensitive photoemission measurement is carried out for 
Cd$_{2}$Re$_{2}$O$_{7}$.

\section{Experimantal}
The synchrotron radiation experiments were performed at beam line BL25SU in 
SPring-8 with using a Scienta SES-200 analyzer. The excitation 
photon energies were selected as 600$\sim $1100 eV for the bulk-sensitive 
measurements. The overall energy resolution was set to about 80 meV for the 
valence band region and 200 meV for the core level photoemission. The 
measurements were done at the sample temperatures of 20 and 250 K below 
and above the transition temperature of 200 K. Clean sample surfaces were 
obtained by cleaving the single-crystal samples \textit{in situ} under an ultra-high vacuum 
of better than 5$\times $10$^{ - 10}$ Torr.

\section{Results and discussions}
Figure 1 shows the valence band spectrum at 20 K measured at the photon 
energy of 600 eV. A finite DOS with a sharp peak exists at the Fermi level. 
The deep valley structure in 1$\sim $3 eV is clearly seen in agreement with 
the band calculation~\cite{cit6}. Considering the cross-sections at the photon 
energy of 600 eV, the Re 5d states are dominant in the peak structure. 
According to a band calculation, the gap structure between 1 and 3 eV is 
derived from the hybridizing between the Re 5d and O 2p orbits. The valence 
band is separated by this gap as the bonding-state near 4 eV and the 
anti-bonding state near the Fermi level. Meanwhile the peaks at 11 and 22 eV 
correspond to the Cd 4d and O 2s states.

\begin{figure}[!ht]
\begin{center}
\includegraphics[width=0.7\textwidth]{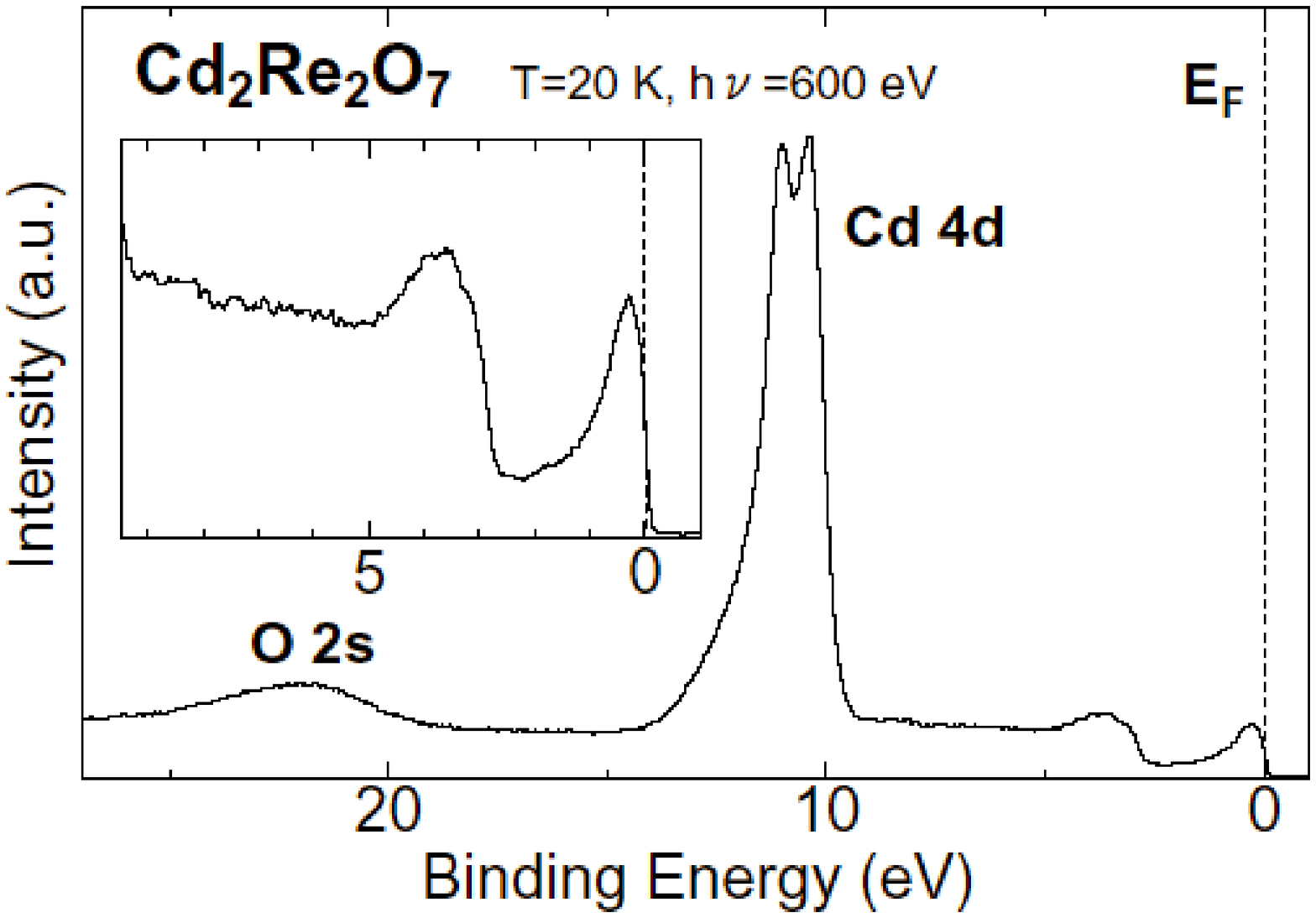}
\end{center}
\caption{Valence band spectrum of Cd$_{2}$Re$_{2}$O$_{7}$ pyrochlore at 20K measured at the photon energy of 600 eV.}
\label{Fig1}
\end{figure}

Figure 2 shows the difference of photoemission spectra near the Fermi level 
below and above 200 K. The spectral intensities of both 20 and 250 K spectra 
are tentatively normalized by the intensity of the localized O 2s inner core 
state. The normalization in the region above the binding energy of E$_{B}$=1 
eV provides the equivalent result. As shown in the inset, the DOS at the 
Fermi level is larger at 20 K than at 250 K. It is consistent with the 
behavior of the electric resistivity~\cite{cit1,cit2}. Furthermore, the spectral shape 
of the peak is obviously different between 20 and 250 K. The spectrum at 20 
K has a broad and rather flat top. This change may be ascribed to the 
removal of the degeneracy of the Re 5d orbital constituting the peak below 
200 K as the result of the lowering of the inversion symmetry of the 
B-O$_{6}$ octahedra.

\begin{figure}[!ht]
\begin{center}
\includegraphics[width=0.4\textwidth]{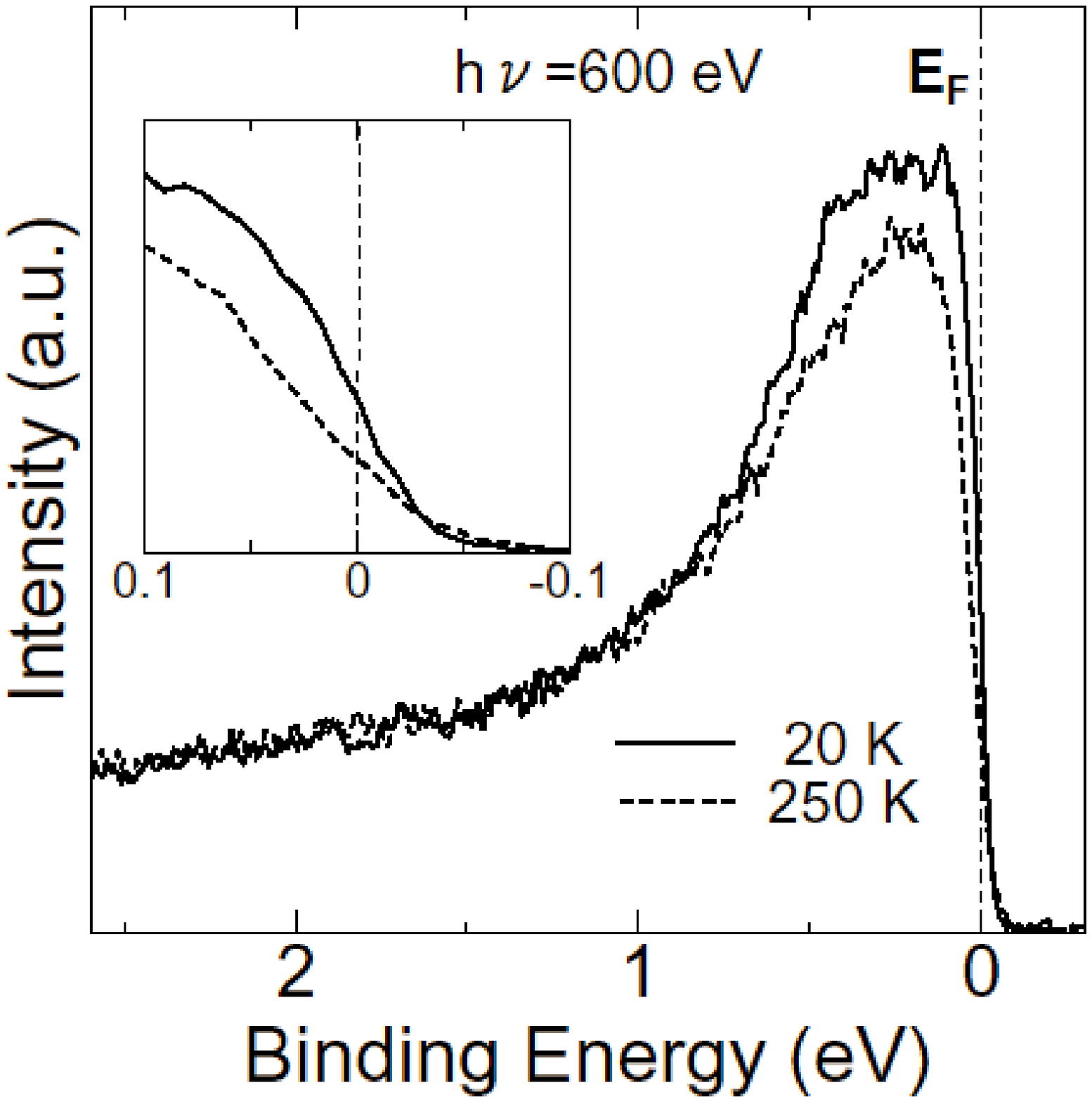}
\end{center}
\caption{Temperature variation of the high-resolution photoemission spectra near E$_{F}$. These spectra are normalized by the intensity of the O 2s core state.}
\label{Fig2}
\end{figure}

As a typical inner core state, we show the temperature dependence of the Re 
4f spectra in Fig. 3(a). The peaks at 42 and 45 eV correspond to j=7/2 and 
5/2 states split by the spin-orbital interaction. A shoulder structure is 
also seen on the higher binding energy side of the j=5/2 component. This 
shoulder is understood as a bulk component because its intensity shows very 
poor h$\nu $ dependence. With decreasing the temperature, the spectral 
weight transfers from the shoulder to the peak. It is recognized that the 
intensity of the j=7/2 component increases also at 20 K whereas the 
intensity in the valley region decreases. According to a deconvolution, 
another shoulder associated with the j=7/2 is hidden in the valley region. 
The origin of these shoulder structures is not fully known yet, but this 
temperature change in the spectral weight may correspond to the spectral 
change near the Fermi level dominated by the Re 5d partial density of state 
(PDOS). One possible scenario is to interpret these shoulders as 
charge-transfer (C-T) satellites. The reduction of C-T satellite at 20 K 
means a reduction of hybridization between the Re 5d and conduction band 
states. Then the Re 5d component in the state near the Fermi level will 
increase and provide higher intensity at 20 K. X-ray BIS study will help to 
check this scenario in near future.

\begin{figure}[!ht]
\begin{center}
\includegraphics[width=0.9\textwidth]{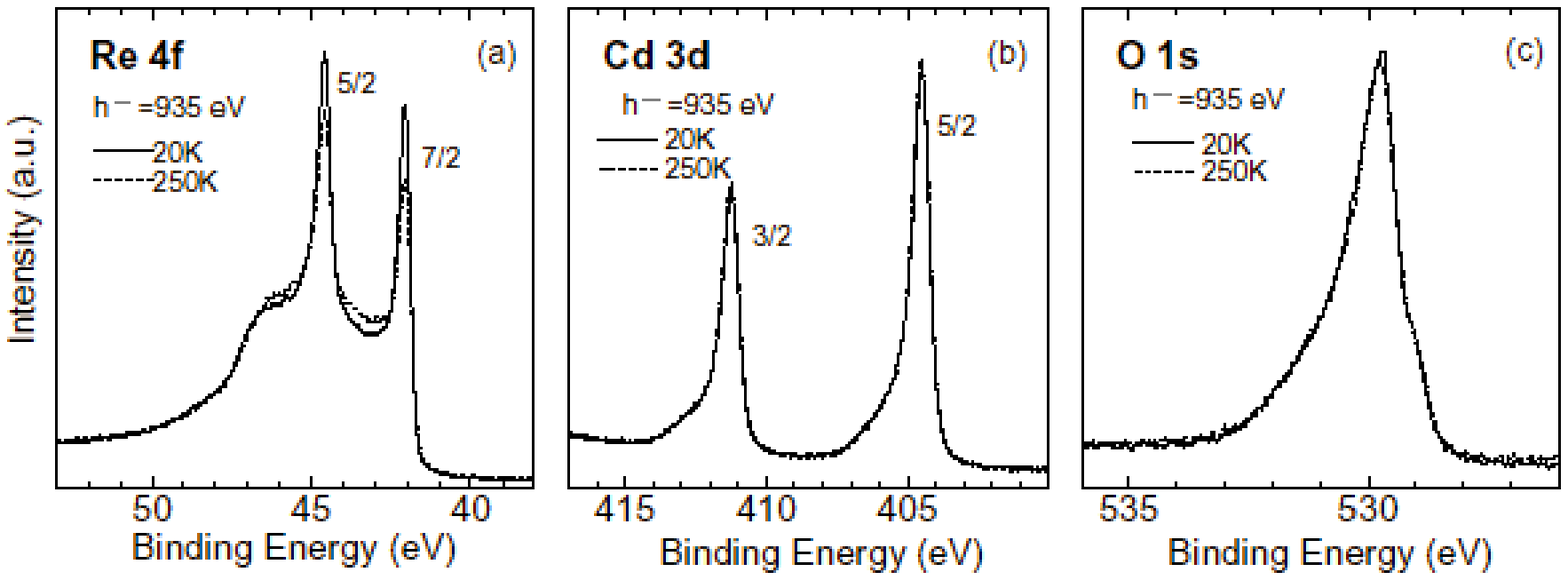}
\end{center}
\caption{Photoemission spectra of (a) Re 4f, (b) Cd 3d, and (c) O 1s inner cores at 20 and 250 K at the photon energy of 935 eV.}
\label{Fig3}
\end{figure}

On the contrary, both Cd 3d and O 1s spectra in Fig. 3(b) and (c) show very 
small change between 20 and 250 K. Beside sharp main peaks small shoulders 
are recognized at higher binding energies. From the photon energy dependence, 
we ascribe these shoulders to surface components. In the O 1s spectrum, the 
shoulder at lower binding energy is not attributed to a surface component 
but may be derived from the existence of two different crystallographic 
sites in the oxygen.

\section{Conclusion}
The temperature dependence of the bulk-sensitive photoemission spectra are 
investigated in Cd$_{2}$Re$_{2}$O$_{7}$ by using soft X-ray synchrotron 
radiation. The valence band spectrum agrees qualitatively well with the 
recent band calculation. The temperature dependence of the Re 4f inner core 
spectra exhibits the dominance of the Re 5d states at the Fermi level. The 
characteristic change in the peak structure near the Fermi level suggests 
the change of band structure dominated by the Re 5d states across 200 K.

\section{Acknowledgement}
We are grateful to Dr. T. Muro and Dr. Y. Saitoh for their experimental 
assistance. The research was performed at SPring-8 under the support of a 
Grant-in-Aid for the COE Research (10CE2004) of the Ministry of Education, 
Culture, Sports, Science, and Technology (MEXT), Japan.

\end{document}